\documentclass[aps,showpacs,twocolumn,amsmath,amssymb,prl]{revtex4-1} 

\usepackage{graphicx} 
\usepackage{bm}

\begin{document} 
 
\title{Lifetime statistics of quantum chaos studied by a multiscale analysis} 
 
\author{A. Di Falco$^1$, T. F. Krauss$^1$ and A. Fratalocchi$^{2}$} 

\email{andrea.fratalocchi@kaust.edu.sa} 
\homepage{www.primalight.org}

\affiliation{
$^1$School of Physics and Astronomy, University of St. Andrews, North Haugh, St. Andrews KY16 9SS, UK  \\
$^2$PRIMALIGHT, Faculty of Electrical Engineering; Applied Mathematics and Computational Science, King Abdullah University of Science and Technology (KAUST), Thuwal 23955-6900, Saudi Arabia
}

\date{\today} 
 
\begin{abstract} 
In a series of pump and probe experiments, we study the lifetime statistics of  a quantum chaotic resonator when the number of open channels is greater than one. Our design embeds a stadium billiard into a two dimensional photonic crystal realized on a Silicon-on-insulator substrate. We calculate resonances through a multiscale procedure that combines graph theory, energy landscape analysis and wavelet transforms. Experimental data is found to follow the universal predictions arising from random matrix theory with an excellent level of agreement.
\end{abstract} 
 
 
\maketitle 

\paragraph{Introduction. ---}
In a seminal paper of 1917, Albert Einstein provided a remarkable insight about the failure of energy quantization schemes when applied to nonintegrable systems and raised the question \emph{"does chaos lurk in the smooth, wavelike, quantum world?"} \cite{kox96:_collec_paper_of_alber_einst_volum,gutzwiller92:_quant_chaos}. This problem was practically ignored until 1970, when theoretical physicists led by M. Gutzwiller investigated the implications of classical chaos for semiclassical quantum systems and founded the research field known as quantum chaos \cite{OTT}. Besides quantum mechanics, the problem posed by Einstein is of great importance due to the large number of systems that can be treated semiclassically \cite{makarov09:_ray_and_wave_chaos_in_ocean_acous, QCAI}. Among them, electromagnetic waves have stirred particular interest. In electrodynamics, quantum chaos originates from  the isomorphisms between Schr\"odinger and Maxwell's equations in two spatial dimensions and is manifested in dielectric optical microresonators whose forms mimic classically chaotic billiards. In these geometries, semiclassical methods and Random Matrix Theory (RMT) provide theoretical predictions that have found to well agree with several experiments, leading not only to fundamental discoveries but also to novel concepts in the field of laser devices \cite{PhysRevLett.94.233901,noeckel97:_ray_and_wave_chaos_in,PhysRevLett.90.063901,Gmachl05061998,PhysRevLett.99.224101,PhysRevLett.101.084101,PhysRevLett.102.044101}. One of the most fundamental aspects of quantum chaos lies in the universality of its eigenmodes, first conjectured by Bohigas, Giannoni and Schmit in 1984 for closed systems \cite{PhysRevLett.52.1} and then theoretically extended to open media through random matrix theory \cite{haake01:_quant_signat_chaos,PhysRevA.79.053806}.  According to RMT, the spectral resonances of a quantum chaotic system show universal probability distributions, which depend only on the symmetries of the original equation and the number of open channels $\mathcal{N}$ considered \cite{Guhr1998189,0305-4470-32-5-003}. However, while there exists a large experimental literature on the study of the position of each resonance (i.e., the real part of the eigenvalue of an eigenmode), experimental work on the lifetime statistics (i.e., the imaginary part, or equivalently, the resonance width) is still at the beginning. The measurement of the lifetimes, in fact, is an extremely challenging task due to resonances overlapping in the eigenmodes spectrum. In this field of research, the present literature consists only of two experiments, which have been performed on microwave billiards characterized by a single open channel \cite{PhysRevLett.100.254101,PhysRevLett.74.62}. Lifetime statistics extracted from measurements have reported to significantly deviate from the prediction of RMT for $\mathcal{N}=1$, showing a behavior whose physical origin is still not understood \cite{PhysRevLett.100.254101}. Besides that, work in correspondence of $\mathcal{N}>1$ (where the form of the statistics is strongly different from the case $\mathcal{N}=1$) is missing, as well as an experimental analysis at optical wavelengths, which are extremely interesting due to the possibility of exploiting nonlinear effects to control quantum chaos phenomena (see, e.g., \cite{PhysRevLett.102.044101,Gmachl05061998}).\\
\begin{figure}
\centering
\includegraphics[width=6.5cm]{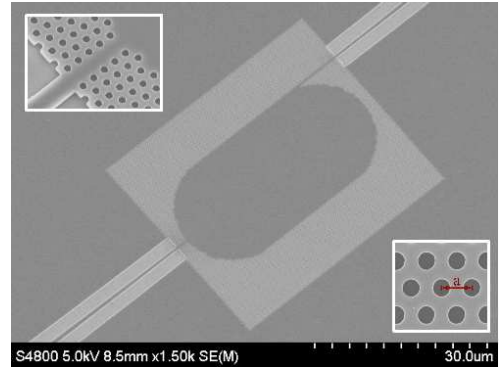}
\caption{Color Online. SEM image of the open stadium billiard realized within a two dimensional photonic crystal. 
\label{sample}
}
\end{figure}
In this Letter, we study the lifetime statistic of a chaotic optical microresonator in the case of $\mathcal{N}>1$. In a series of pump and probe experiments, we injected a broadband source signal into a stadium microresonator and collected the output by a fiber spectrograph. Resonance lifetimes have then been extracted from the power spectrum by a multiscale procedure which combines wavelet transforms and energy landscape analysis \cite{jaffard01:_wavel,wales03:_energ_lands}. We performed an experimental campaign on five different samples, and analyzed more than $700$ resonances. Experimentally measured statistics were found to perfectly match the universal predictions of RMT. 
\begin{figure}
\centering
\includegraphics[width=7.5cm]{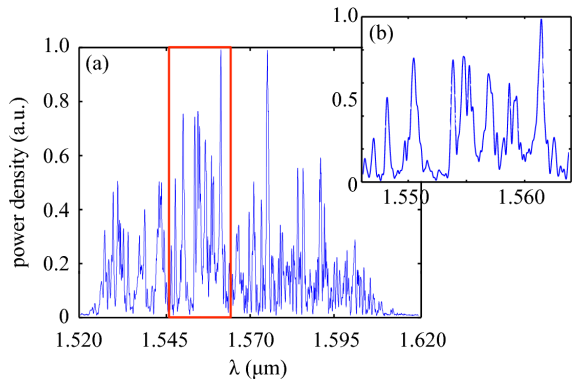}
\caption{Color Online. (a) Experimental power density spectrum $\mathcal{P}(\lambda)$ retrieved for a sample with $f=2.88$; (b) enlarged portion (a) for $\lambda\in[1.546 \mu\mathrm{m},1.564 \mu\mathrm{m}]$.
\label{spec}
}
\end{figure}
\paragraph{Sample \& experimental setup. ---}
Experiments on wave chaos have been traditionally performed in the microwave regime due to possibility of using large geometries ($\approx mm$) and low loss metals acting as mirrors. At optical wavelengths, conversely, metals are lossy and low absorption dielectrics should be employed. However, a naive implementation of a chaotic billiard with dielectrics results into large fresnel losses, and RMT statistics would no longer be universal \cite{PhysRevA.77.013834}. In order to overcome this problem ad design an optical cavity that mimics a microwave billiard, we resorted to Photonic Crystals (PhC) technology \cite{SakodaBook,PhysRevLett.79.4147}. In particular, we embedded a fully chaotic stadium geometry \cite{chernov06:_chaot_billiard} in a two dimensional photonic crystal made by a triangular lattice of cylindrical air holes in silicon (Fig. \ref{sample}). The sample was fabricated on a Silicon on Insulator substrate, with a SiO$_2$ buffer layer of  $2 \mu m$ and a Silicon capping layer of $220 nm$. The PhC pattern was written with a standard lithographic procedure, involving exposure with an e-beam system, development of the photoresist and transferring of the pattern on to the substrate with reactive ion etching step (see e.g. \cite{difalcoapl}).  The confinement in the transverse direction is guaranteed by total internal reflection, whereas the PhC cavity ensures light confinement in the plane. The sample was designed to work with the electric field parallel to the surface (TE modes). Inside a photonic crystal, wavelengths $\lambda$ that fall in the PhC gap $\Delta\lambda$ cannot propagate due to a zero density of electromagnetic states \cite{SakodaBook}. In such a frequency range, light energy entering into the billiard gets totally reflected by the PhC walls and escapes in the input and output waveguides. Other sources of outcoupling losses, besides the waveguides, may be identified in the Silicon-air and Silicon-SiO$_2$ interfaces, owing to the finite dielectric permittivity of the Silicon, and in the PhC itself. The latter, due to its finite extension, is expected to sustain a finite ---albeit evanescently small--- energy leakage in the plane. All these five elements (input/output waveguides, PhC and upper/lower interfaces) are completely independent and work as $\mathcal{N}=5$ autonomous open channels.\\In order to collect a sufficiently large statistics, we designed different samples with the same billiard area (of $700\mu\mathrm{m}^2$) and diverse filling factor $f=\frac{r}{a}$, being $r$ the radius of the single air hole and $a$ the lattice spacing (Fig. \ref{sample}). By varying the filling factor in the range $f\in[0.25,0.3]$, we guaranteed a photonic bandgap for electromagnetic TE modes of $\Delta\lambda> 400$ nm, which is sufficiently larger than the bandwidth of the laser source employed for the experiments. The experimental setup consisted of a polarized C+L band amplified spontaneous emission source (central wavelength 1575 nm, bandwidth 110 nm), which injected light into a single mode fiber that was coupled to the sample by a $60$x aspheric lens, with antireflection coating. Light emerging from the output channel was then collimated by a $40$x aspheric lens into a second single mode fiber, split between a photodetector (to monitor the coupling optimization) and an ANDO optical spectrum analyzer. All spectra were acquired with a resolution of 10 pm.

\begin{figure}
\centering
\includegraphics[width=8.5cm]{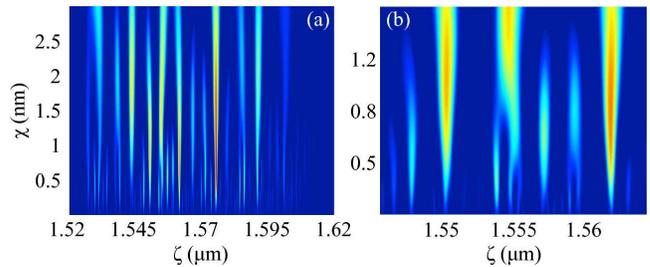}
\caption{Color Online. (a), (b): continuous wavelet transform of the power density spectrum $\mathcal{P}(\lambda)$ of Fig. \ref{spec}a and Fig. \ref{spec}b, respectively.
\label{wvt}
}
\end{figure}

\begin{figure}
\centering
\includegraphics[width=8.5cm]{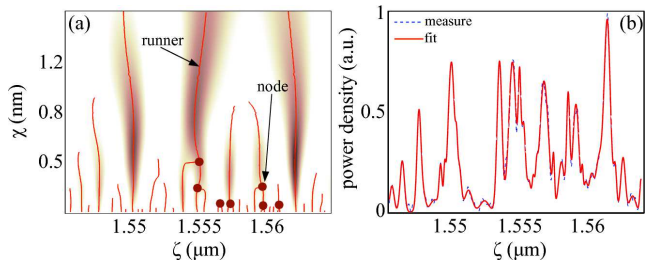}
\caption{Color Online. (a) uphill pathways analysis (runners as solid lines and nodes as circle markers) of the wavelet transform of Fig. \ref{wvt}b, reported for reference as pseudocolor plot; (b) theoretical reconstructed spectrum versus experimental measure.  \label{run}
}
\end{figure}

\paragraph{Multiscale analysis of the resonances. ---} 
Figure (\ref{spec}) shows a typical power density spectrum $\mathcal{P}(\lambda)$ obtained from a single measurement. 
We observe strong clustering effects, with several resonances overlapping in the spectrum (see Fig. \ref{spec}b). We emphasize that, due to such overlapping, no standard method of resonances extraction produced satisfactory results. For this reason, we developed an original approach that combines ideas from multiresolution analysis and graph theory. More specifically, we begin by applying a continuous wavelet transform $(\mathcal{W}\;\mathcal{P})(\zeta,s)$ to the power density spectrum $\mathcal{P}$:
\begin{equation}
\label{wav}
(\mathcal{W}\;\mathcal{P})(\zeta,\chi)=\int_{-\infty}^\infty \mathcal{P}(\lambda)\frac{1}{\sqrt{\chi}}\psi\bigg(\frac{\lambda-\zeta}{\chi}\bigg)\mathrm{d\lambda},
\end{equation}
being $\chi$, $\zeta$ scaling and translation parameters, respectively, and $\psi$ the so-called mother wavelet, which is a compact function in $L^2(\mathbf{R})$ possessing a zero mean value and satisfying $\int_{-\infty}^\infty\frac{|\tilde\psi(f)|^2}{|f|}\mathrm{df}<\infty$ ($\tilde{\psi}$ denotes the Fourier transform of $\psi$). To extract Lorentzian-like linewidths out of the spectrum, we choose the following symmetric wavelet $\psi=(2-4t^2)e^{-t^2}$, which is defined from the second derivative of a Gaussian function.
%\begin{equation}
%\label{gaus2}
%\psi(t)=\bigg(\frac{2}{\pi}\bigg)^{\frac{1}{4}}\frac{2-4t^2}{\sqrt{3}}e^{-t^2},
%\psi(t)=(2-4t^2)e^{-t^2},
%\end{equation}
%with $c=(2/\pi)^{\frac{1}{4}}/\sqrt{3}$normalized to have a unit energy $\int_{-\infty}^\infty|\psi(t)|^2\mathrm{dt}=1$.
The integral transform (\ref{wav}), when applied to the power density spectrum resulting from a measurement, provides a geometric visualization of the inner structure of the spectral resonances ---given by the regions of high density in Fig. \ref{wvt}--- which change according to the scale $\chi$ considered. As $\chi$ increases, resonances found at smaller scales evolve into tree-like structures  and eventually join together (Fig. \ref{wvt}b). The topology of the surface $(\mathcal{W}\;\mathcal{P})$ encodes the structure of all the resonances of the system. Isolated resonances, in particular, are observed due to the missing of any tree structure in $(\mathcal{W}\;\mathcal{P})$, while a link between high-density regions is the signature of the presence of a cluster. We describe clusters and isolated resonances by an uphill landscape analysis \cite{wales03:_energ_lands} applied to the wavelet transform. Figure \ref{run}a illustrates this analysis when applied to the portion of the spectrum of Fig. \ref{spec}b. In particular, we employ a series of runners (Fig. \ref{run}a, solid lines), which evolve along $\chi$ following the curve of minimum steepness on the surface $(\mathcal{W}\;\mathcal{P})$, and nodes (Fig. \ref{run}a, circle dots) denoting the intersection among different runners. 
We geometrically describe tree-like structures among paths of different runners by building an adjacency matrix $\mathcal{A}$, with $\mathcal{A}_{ij}=1$ if the path of runners $i$ and $j$ overlaps, and $\mathcal{A}_{ij}=0$ otherwise. The undirected simple graph $\mathcal{G}$, arising from $\mathcal{A}$, carries the information on both the structure and the composition of all the clusters of resonances in the spectrum.  In conjunction with the wavelet transform $(\mathcal{W}\;\mathcal{P})$, it provides the required knowledge to extract position and spectral linewidth of the billiard eigenmodes. Isolated resonances, characterized by $A_{ij}=0$ for $j\neq i$, are described by following the path of the corresponding runner up to the first maximum in the surface $(\mathcal{W}\;\mathcal{P})$, which yields the highest overlap according to the $L^2$ distance defined by (\ref{wav}). At the maximum point in the plane $(\chi,\zeta)$, the Full Width Half Maximum (FWHM) linewidth $\delta\lambda$ is $\sqrt{2\log{2}}$ times the waist of the gaussian part of the wavelet, and reads $\delta\lambda=\chi\sqrt{\log{2}/2}$, while the resonance position $\lambda_0$ is given by $\lambda_0=\zeta$. In the presence of resonance clusters, which show nonzero connections in the adjacency matrix (i.e., $A_{ij}\neq 0$ for $j\neq i$), we found optimal $(\lambda_0,\delta\lambda)$ by a nonlinear simplex optimization search in the subspace spanned by the runners path in the plane $(\chi,\zeta)$ of the surface $(\mathcal{W}\;\mathcal{P})$. Stemming from the theoretical parameters $(\lambda_0,\delta\lambda)$ obtained through the combined analysis on $(\mathcal{W}\;\mathcal{P})$ and $\mathcal{G}$, we finally apply a global nonlinear least-square optimization across the entire spectrum for further increasing the fidelity of our estimated results.
\begin{figure}
\centering
\includegraphics[width=8.5cm]{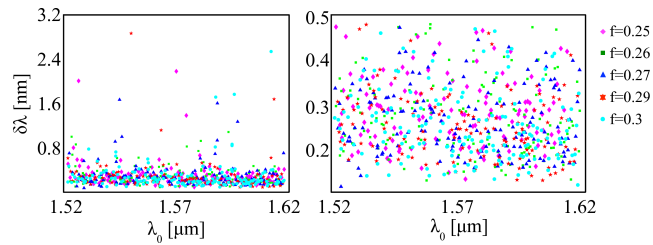}
\caption{Color Online. (a)-(b) Position $\lambda_0$ and FWHM linewidth $\delta\lambda$ of chaotic resonances extracted by the multiscale analysis. (b) shows a magnified version of (a) in the region where the resonances exhibit a random behavior.
\label{data0}
}
\end{figure}
The outcome of this procedure are quite satisfactory. Figure \ref{run}b compares a portion of the theoretical spectrum reconstructed by the method described above and the experimental measure. As seen, the experimental data is barely visible due to an excellent agreement with the fitted curve obtained through the multiscale analysis. We highlight that all experimental spectra (not shown here), were reconstructed with the same degree of precision.

\paragraph{Discussion of results. ---} Figure \ref{data0} reports the distribution of the resonances in the plane $(\lambda_0,\delta\lambda)$ for the five samples considered in our experiments. The figure shows a clear separation of the resonance dynamics into two different regions. A few, isolated short living resonances, and many long living modes with positions that are completely random in $(\delta\lambda,\lambda_0)$. The former, associated to billiard's bouncing-ball eigenmodes corresponding to nodal lines along horizontal and vertical spatial directions, are normally ignored in the statistical analysis of the spectrum \cite{PhysRevA.37.3067}. We therefore restrict our attention to long-living resonances, and study their statistics by applying a cutoff rate of $\delta\lambda\le 0.6$. In order to compare our  results to the corresponding quantum-mechanical predictions, we begin by expressing the scattering matrix $S(\mathcal{E})$ \cite{annphys_220_159,PhysRevA.67.013805}, which relates the vector of incoming $\mathbf{B}$ and outgoing $\mathbf{A}$ electromagnetic field amplitudes by $\mathbf{B}=S\mathbf{A}$, with $S(\mathcal{E})=1-2\pi i\mathcal{W(\mathcal{E})^\dagger}\mathcal{D(\mathcal{E})}^{-1}\mathcal{W(\mathcal{E})}$,
%\begin{align}
%\label{sca}
%S(\mathcal{E})=1-2\pi i\mathcal{W(\mathcal{E})^\dagger}\mathcal{D(\mathcal{E})}^{-1}\mathcal{W(\mathcal{E})},
%\end{align}
$\mathcal{E}$ playing the role of a quantum-mechanical energy, $\mathcal{W}$ the operator modeling the coupling with the environment (i.e., the channel space) and 
%\begin{equation}
$\mathcal{D}=\mathcal{E}-\mathcal{H}_0+i\pi\mathcal{W}\mathcal{W}^\dagger$,
%\end{equation}
being $\mathcal{H}_0=\nabla^2+k^2$ the Hermitian Hamiltonian corresponding to the close billiard. Resonances of the system originate from the poles of the scattering matrix $S$, and therefore, from the eigenvalues $\mathcal{E}=\nu-i\gamma$ of $\mathcal{H}_0-i\pi\mathcal{W}\mathcal{W}^\dagger$. The eigenvalues imaginary part $\gamma$, which corresponds to the distribution of the resonances width of the open chaotic system, can be related to the electromagnetic measured data $(\lambda_0,\delta\lambda)$ from the relation $k^2=\frac{\omega^2}{c^2}=\mathcal{E}$, and reads  $\gamma=\delta\lambda/\lambda_0^3$ (omitting inessential proportionality constants). Probability distributions of eigenvalues $\mathcal{E}$ arising from Hamiltonians $\mathcal{H}_0-i\pi\mathcal{W}\mathcal{W}^\dagger$ have been studied in great detail and reviewed by Fyodorov and Sommers in \cite{10.1063_1.531919}. Closed form expression for the probability of decaying constants $p(\gamma)$ can be obtained by the supersymmetry method and in the limit of weak losses is expressed by the following universal distribution:
\begin{equation}
\label{univ}
p(\gamma)=\frac{g^\mathcal{N}}{\Gamma(\mathcal{N})}\gamma^{\mathcal{N}-1}e^{-g\gamma},
\end{equation}
with $g=(\mathcal{N}\langle \gamma \rangle)^{-1}$. The distribution (\ref{univ}) is uniquely determined by the number of channels $\mathcal{N}$, which defines the dimensionality of the channel space. 
\begin{figure}
\centering
\includegraphics[width=5.5cm]{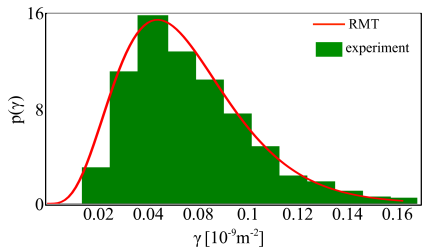}
\caption{Color Online. (histogram) Distribution of resonance imaginary part $p(\gamma)$ and theoretical prediction arising from RMT (solid line). 
\label{data1}
}
\end{figure}
When $\mathcal{N}=1$, Eq. (\ref{univ}) collapses to a simple exponential distribution, while for $\mathcal{N}>1$ it acquires a radically different shape characterized by a nonzero probability maximum.  Figure \ref{data1} compares the experimentally measured distribution of $\gamma$ with the theoretical prediction of RMT with $\mathcal{N}=5$. As seen, an excellent level of agreement is found between theory and experiments. It is worthwhile remarking that we did not employ any fitting parameter in Fig. \ref{data1}, but just the universal distribution (\ref{univ}) with $\mathcal{N}=5$.

\paragraph{Conclusions. ---} Motivated by a question posed by Einstein in the last century, we investigated the universal behavior of quantum systems in the presence of deterministic chaos. By exploiting an isomorphism with Maxwell and Schr\"odinger equations, we designed and experimentally characterized photonic crystal resonators that are fully equivalent to open quantum billiards. Our experimental results, analyzed within a theoretical framework that combines ideas from multiresolution analysis, unambiguously demonstrate the existence of universal statistics in the lifetime of quantum chaos. Universality is a concept of utmost importance in every branch of physics, and its implications at the quantum scale are fundamental not only to add a piece to the puzzle initiated by Einstein's intuition, but also to give rise to new quantum devices that exploit universal phenomena and operate regardless of their microscopic details. 

%\bibliography{../../refbib} 
%merlin.mbs apsrev4-1.bst 2010-07-25 4.21a (PWD, AO, DPC) hacked
%Control: key (0)
%Control: author (8) initials jnrlst
%Control: editor formatted (1) identically to author
%Control: production of article title (-1) disabled
%Control: page (0) single
%Control: year (1) truncated
%Control: production of eprint (0) enabled
%

\end{document}